\documentclass[aps,amsmath,onecolumn, showpacs,superscriptaddress,pre,10.5pt,nofootinbib]{revtex4-1}
\usepackage{amssymb}
\usepackage{amsmath}
\usepackage{graphicx}
\usepackage{array}
\usepackage{physics}
\usepackage{mathtools}
\usepackage{amssymb}
\usepackage{verbatim}
\usepackage{color}
\usepackage{bm}
\usepackage{bbm}
\usepackage{amsthm}
\usepackage[normalem]{ulem}
\usepackage[titletoc,title]{appendix}

\usepackage{fixmath}
\newcommand{\vect}[1]{\mathbold {#1}} 

\usepackage{color}
\usepackage[dvipsnames]{xcolor}
\definecolor{Blue}{rgb}{0.00, 0.00, 1.00}
\definecolor{Red}{rgb}{1.00, 0.00, 0.00}
\definecolor{Green}{rgb}{0.00, 0.50, 0.00}

\usepackage[colorlinks=true, urlcolor=blue, anchorcolor=blue, citecolor=blue,filecolor=blue,linkcolor=blue,menucolor=blue]{hyperref}

\newcommand{\bea}{\begin{eqnarray}}
\newcommand{\eea}{\end{eqnarray}}
\newcommand{\be}{\begin{equation}}
\newcommand{\ee}{\end{equation}}
\newcommand{\nn}{\nonumber}
\newcommand{\bee}{\begin{equation*}}
\newcommand{\eee}{\end{equation*}}

\colorlet{Mycolor1}{green!10!orange!90!}

\def\XXint#1#2#3{{\setbox0=\hbox{$#1{#2#3}{\int}$}
     \vcenter{\hbox{$#2#3$}}\kern-.5\wd0}}

\begin{document}
\title{Anomalous scalings of fluctuations of the area swept by a Brownian particle trapped in a $|x|$ potential}

\author{Naftali R. Smith}
\email{naftalismith@gmail.com}

\affiliation{Department of Environmental Physics,
  Blaustein Institutes for Desert Research, Ben-Gurion University of
  the Negev, Sede Boqer Campus, 8499000, Israel}

\begin{abstract}

We study the fluctuations of the area $A=\int_0^T x(t) dt$ under a one-dimensional Brownian motion $x(t)$ in a trapping potential $\sim |x|$, at long times $T\to\infty$. We find that typical fluctuations of $A$ follow a Gaussian distribution with a variance that grows linearly in time (at large $T$), as do all higher cumulants of the distribution.
However, large deviations of $A$ are not described by the ``usual'' scaling (i.e., the large deviations principle), and are instead described by two different anomalous scaling behaviors:
Moderately-large deviations of $A$, obey the anomalous scaling 
$P\left(A;T\right)\sim e^{-T^{1/3}f\left(A/T^{2/3}\right)}$
 while very large deviations behave as 
 $P\left(A;T\right)\sim e^{-T\Psi\left(A/T^{2}\right)}$.
We find the associated rate functions $f$ and $\Psi$ exactly. 
Each of the two functions contains a singularity, which we interpret as dynamical phase transitions of the first and third order, respectively.
We uncover the origin of these striking behaviors by characterizing the most likely scenario(s) for the system to reach a given atypical value of $A$.
We extend our analysis by studying the absolute area $B=\int_0^T|x(t)| dt$ and also by generalizing to higher spatial dimension, focusing on the particular case of three dimensions.

\end{abstract}

\maketitle

\section{Introduction}

\subsection{Background}

Fluctuations in stochastic, dynamical systems are of fundamental importance in statistical mechanics.
One of the central paradigms over the recent few decades to probe dynamical effects on such fluctuations  has been the study of what is known as ``dynamical'' observables. These are
observables which involve not only the state of the system at a given time, but the cumulative effect of its 
state over a long time window \cite{hugo2009}. 
Such observables cannot be studied just by using standard tools of thermodynamics in equilibrium. Indeed, nonequilibrium behaviors often occur even given that the underlying dynamical process is in thermal equilibrium. In particular, fluctuations of such observables have no description in terms of a Boltzmann distribution. In their full distribution (including large deviations \cite{O1989, DZ, Hollander, hugo2009, Touchette2018, Jack20}), 
a number of striking effects may occur, e.g., dynamical phase transitions (DPTs) \cite{exclusion, exclusion1, kafri, glass, NemotoEtAl19, CVC23, Baek15, Baek17, Baek18, SmithMeerson2019, MeersonSmith2019, TouchetteMinimalModel, Touchetteoneparticle, MukherjeeSmith23OF, MukherjeeSmith23CH}: Singularities of large-deviation functions which describe the fluctuations in the long time limit. DPTs are analogous to phase transitions in systems that are in equilibrium. 


Consider a stochastic, dynamical process $x(t)$ in continuous time $t$.
Dynamical observables (of a relatively simple class) are given by long-time averages
\be
\label{DynamicalObservable}
A=\int_{0}^{T}u\left(x\left(t\right)\right)dt \, ,
\ee
where $u(x)$ is a given function.
In the long time limit $T\to\infty$, fluctuations of dynamical observables often obey a large-deviations principle (LDP),
\be
\label{LDP}
P\left(A;T\right)\sim e^{-TI\left(A/T\right)} \, .
\ee
The function $I(a)$ is known as the rate function. It is generically convex, and vanishes at the ensemble-average value of $a$.
Donsker-Varadhan (DV) theory provides a method to calculate $I(a)$, by solving the auxiliary problem of finding the largest eigenvalue of a ``tilted'' (i.e., modified) generator of the dynamics of $x(t)$. The latter problem also yields the coefficients of the linear growth of the cumulants of $A$ with $T$ in the limit $T\to\infty$.
However, over recent years, DV theory has been found to break down in several cases  \cite{HT09, NMV10, NT18, MeersonGaussian19, GM19, DH2019, JackHarris20, BH20, BKLP20, MLMS21, MGM21, GIL21, GIL21b, Smith22OU, NT22, SmithMajumdar22}, and with it also the LDP \eqref{LDP}. 
 One cause for the possible failure of DV theory is that the spectrum of the tilted generator may not bounded from above \cite{NT18, DH2019, BH20, Smith22OU, NT22}. 
Instead of the LDP \eqref{LDP}, one may then observe anomalous scalings of the type
$P\left(A;T\right)\sim e^{-T^{\alpha}I\left(A/T^{\beta}\right)}$
with exponents $\alpha$ and $\beta$ that are `anomalous' in the sense that they differ from the standard ones $\alpha=\beta=1$.

In this paper, we consider the dynamics of the position $x(t)$ of a Brownian particle trapped by an external potential $\sim|x|$.
Dynamical properties of this process have been studied extensively. In particular, its exact propagator was found in \cite{TSJ10}, and from it the stationary two-time position correlation function was obtained. The propagator was also reproduced, in the weak-noise regime, in \cite{CBHJ13} (see also the recent work \cite{DBK23} for an extension to potentials that are given by $\sim|x|^\sigma$ with $0<\sigma<1$).

We consider fluctuations of the area 
$A=\int_{0}^{T}x\left(t\right)dt$
 swept by the particle, corresponding to $u(x)=x$ in Eq.~\eqref{DynamicalObservable} above (so that $A/T$ is the time-averaged position of the particle). One of the physical motivations for studying this problem (and/or its extensions to higher spatial dimensions) can be obtained by interpreting $x(t)$ as the velocity of a particle. The deterministic component of the force acting on this particle may arise due to dry friction \cite{TSJ10} or due to damping in the ultrarelativistic limit \cite{DMR97}, and $A$ becomes the position of the particle at time $T$.
In the remainder of the paper, however, we interpret $x(t)$ as the position of the particle (as a function of time $t$) and refer to it as such.

The closely-related problem of the long-time average position of a Brownian particle with constant external drift which is directed towards a reflecting wall was studied in Refs. \cite{Fatalov17, BH20} (see also Refs.~\cite{Meyn08, DM10} in which an analogous discrete-time model was studied). It was found that, while the LDP \eqref{LDP} holds in the left tail of the distribution as successfully described by DV theory, the right tail exhibits a sub-exponential decay in time, and DV theory breaks down.
In Ref.~\cite{DM10}, a regime of very large deviations $A \sim T^2$ was studied, and it was found to display anomalous scaling and with a rate function that is non-convex and that exhibits a singularity (i.e., a dynamical phase transition).
However, a full understanding of these distributions that includes all regimes is lacking for both of these models (i.e., Brownian particle in a $\sim|x|$ potential, and reflected Brownian motion with drift).

In this work, we obtain a more complete understanding of the fluctuations of the area $A$ for a Brownian particle in a $\,\sim \! |x|$ potential, using different theoretical tools for the different parts of the distribution. In the typical-fluctuations regime, we apply a perturbative DV approach (which succeeds despite the failure of the non-perturbative one), rather similar to the one used in Ref.~\cite{Smith22OU}. 
 In this system, very large deviations (i.e., $A \sim T^2$) are not described by any form of the DV theory, because such fluctuations require the particle to travel very far from the origin within a (relatively) short period of time. They are instead captured correctly by the ``geometrical optics'' framework 
\cite{GF03, NE11, BHSH18, NT18, Meerson19, AiryDistribution20, SmithMeerson2019, MeersonSmith2019, AG21, NT22, Meerson23RA, BarMeerson2023, MukherjeeSmith23CH}. This framework is essentially a weak-noise theory in which the probability of the rare event is approximated by that of the ``optimal" (most likely) realization of the process that leads to this event (in a more general context this method is also known  by the names optimal fluctuation method, instanton method or weak-noise theory
\cite{Onsager, MSR, Freidlin}).
Finally, the (highly nontrivial) moderate-deviations regime  $A \sim T^{2/3}$, is described by combining the perturbative DV and geometrical optics methods \cite{Smith22OU, SmithMajumdar22}.

\subsection{Model definition, rescaling and summary of main results}

We consider the dynamics of an overdamped Brownian particle trapped by an external potential $V(x) = \mu |x|$ (with $\mu>0$), as described by the Langevin equation
\be
\label{LangevinOrig}
\dot{x}\left(t\right)=-\mu \, \text{sgn}\left(x\left(t\right)\right)+\sqrt{2D}\,\xi\left(t\right) \, .
\ee
Here 
\be
\text{sgn}\left(x\right)=\begin{cases}
1, & x>0,\\[1mm]
0, & x=0,\\[1mm]
-1, & x<0
\end{cases}
\ee
is the sign function, $D$ is the diffusivity of the particle,
and  $\xi(t)$ is a white noise with
$\left\langle \xi\left(t\right)\right\rangle =0$
and
$\left\langle \xi\left(t\right)\xi\left(t'\right)\right\rangle =\delta\left(t-t'\right)$.

The particle starts at the center of the trap, $x(t=0) = 0$.
We are interested in the distribution of the area swept by this particle up to time $T$,
\be
\label{AreaDef}
A=\int_{0}^{T}x\left(t\right)dt \, .
\ee
Note that $A/T$ is the time-averaged position of the particle over the time window $[0,T]$.

Under the rescaling
\be
\mu x/\left(2D\right)\to x,\qquad\mu^{2}t/\left(2D\right)\to t
\ee
of time and space, respectively, Eq.~\eqref{LangevinOrig} becomes
\be
\label{Langevin}
\dot{x}\left(t\right)=-\text{sgn}\left(x\left(t\right)\right)+\xi\left(t\right) \, .
\ee
The rescaled trapping potential is $V(x) = |x|$,
while $A$ and $T$ are rescaled by $4D^{2}/\mu^{3}$ and $2D/\mu^{2}$, respectively. From here one immediately finds that the distribution of $A$ takes the scaling form
\be
\label{RescalingPAT}
\mathcal{P}_{\mu,D}\left(A;T\right)=\frac{4D^{2}}{\mu^{3}}P\left(\frac{\mu^{3}A}{4D^{2}} \, ; \, \frac{T\mu^{2}}{2D}\right)\,,
\ee
where $P(\dots)$ is dimensionless and so are its arguments.
Note that Eq.~\eqref{RescalingPAT} can also be reached by a simple dimensional analysis.
In the rest of the paper, we shall use the rescaled variables (which correspond to the choices $\mu=1$, $D=1/2$) unless explicitly stated otherwise.

From the (exact) mirror symmetry of the problem, one immediately finds that the distribution of $A$ is exactly mirror symmetric too, i.e., $P(A;T) = P(-A;T)$. Moreover, at long times, the distribution of the position of the particle relaxes to its steady-state, Boltzmann distribution
$p_{s}\left(x\right)\propto e^{-2\left|x\right|}$.
$p_{s}\left(x\right)$  is of course also mirror symmetric, with zero mean. It therefore follows from ergodicity that with probability 1, the time average of the particle's position $A/T$ converges, in the large-$T$ limit, to its corresponding ensemble-average value, which is 
$\int_{-\infty}^\infty x \,  p_s(x)dx=0$. However, at large but finite $T$ there will be fluctuations from this expected value, and they are the object of our study.

Let us now summarize our main findings.
We find that typical fluctuations of $A$ are described by a Gaussian distribution whose variance grows, at long times, as $5T/4$. 
Remarkably, we find that, despite the fact that the standard LDP \eqref{LDP} breaks down, all of the cumulants of the distribution grow linearly in time, with coefficients that are given by the power-series expansion of the function $\lambda(k)$ given below in Eq.~\eqref{lambdakSol}. We give the lowest four nonvanishing coefficients explicitly in Eq.~\eqref{kappasols} and verify them numerically, see Fig.~\ref{fig:cumulants}.

Furthermore, We find that there are two nontrivial large-deviation scaling regimes.
Moderately-large fluctuations, of order $A \sim T^{2/3}$, are described by the anomalous LDP
\be
\label{fScaling}
P\left(A;T\right)\sim\exp\left[-T^{1/3}f\left(\frac{A}{T^{2/3}}\right)\right]
\ee
with the rate function
\be
\label{fsol}
f\left(y\right)=\min_{0\le z\le y}\left[\frac{2^{3/2}\sqrt{z}}{3^{1/2}}+\frac{2}{5}\left(y-z\right)^{2}\right] \, ,
\ee
see Fig.~\ref{fig:PA} below and  Fig.~\ref{fig:fofy} in Appendix \ref{app:fofy}.
Remarkably, $f(y)$ is non-convex and exhibits a first-order dynamical phase transition at a critical value $y_{c}={15}^{2/3} \! /2^{4/3}=2.41372\dots$, corresponding to a ``condensation'' phenomenon.
We find that very large fluctuations, of order $A \sim T^2$, coincide in the leading order with those of the related problem studied in \cite{DM10}. They are described by the anomalous LDP
\be
\label{PsiScaling}
P\left(A;T\right)\sim e^{-T\Psi\left(A/T^{2}\right)}
\ee
with the non-convex rate function
\be
\label{PsiSol}
\Psi\left(w\right)=\begin{cases}
\frac{2^{3/2}\sqrt{w}}{3^{1/2}}\,, & 0<w<\frac{1}{6}\,,\\[2mm]
\frac{3}{8}\left(2w+1\right)^{2}\,, & w \ge \frac{1}{6}\,
\end{cases}
\ee
which exhibits a third-order dynamical phase transition at the critical value $w=1/6$, see Fig.~\ref{fig:GeometricalOpticsAndPsi}(b).
Note that the two results \eqref{fScaling} and \eqref{PsiScaling} match smoothly in their joint regime of validity, $T^{2/3} \ll A \ll T^2$.
Likewise, the Gaussian distribution of typical fluctuations matches smoothly with the parabolic behavior of $f(y)$ around $y=0$.

The rest of the manuscript is organized as follows.
In Section \ref{sec:typical} we analyze the regime of typical fluctuations and obtain the linear temporal growth of the cumulants at long times.
In Section \ref{sec:veryLarge} we analyze the very-large-fluctuations regime $A\sim T^2$.
In Section \ref{sec:moderate} we build on the results of the two previous sections to study the moderate-fluctuations regime $A \sim T^{2/3}$.
We briefly outline the extension of some of our results to the absolute area and to higher spatial dimensions in Sections \ref{sec:absoluteArea} and \ref{sec:higherDim}, respectively.
We summarize and discuss our findings in Section \ref{sec:discussion}.
Some technical details are given in the Appendices.

\section{Typical fluctuations}
\label{sec:typical}

In principle, one can calculate the full distribution of $A$ from the propagator $p\left(x,t\,|\,x',t'\right)$ of $x(t)$, which, for the process \eqref{Langevin} was calculated exactly in Ref.~\cite{TSJ10}.
For instance, the variance is given (exactly) by
\be
\label{AfromCdef}
\left\langle A^{2}\right\rangle =\int_{0}^{T}dt\int_{0}^{T}dt'\,\left\langle x\left(t\right)x\left(t'\right)\right\rangle  \, .
\ee
In the limit $T\gg1$ that we focus on in the present work, one approximates
$\left\langle x\left(t\right)x\left(t'\right)\right\rangle \simeq C\left(\left|t-t'\right|\right)$
where
\be
\label{Cdef}
C\left(\tau\right)=\int_{-\infty}^{\infty} dx\int_{-\infty}^{\infty} dx'\,xx'p\left(x,\tau\,|\,x',0\right)p_{s}\left(x'\right)
\ee
is the steady-state two-time correlation function (which was extracted in \cite{TSJ10} from the propagator), $p_{s}\left(x\right)$ being the equilibrium steady-state distribution of $x$.
Plugging this into \eqref{AfromCdef}, and using that $C(\tau)$ decays rapidly with $\tau$ at $\tau \gg 1$, one finds that the variance of the distribution grows linearly in time as
\be
\label{AfromCWithIntegral}
\left\langle A^{2}\right\rangle \simeq T\int_{-\infty}^{\infty}C\left(\left|\tau\right|\right)d\tau=2T\int_{0}^{\infty}C\left(\tau\right)d\tau \, ,
\ee
which is nothing but the Green-Kubo relation \cite{Green54, Kubo57}.
In Appendix \ref{app:correlationFunction} we  substitute the correlation function from \cite{TSJ10}, solve this integral and find that
\be
\label{varVsTDHalf}
\left\langle A^{2}\right\rangle _{c}\simeq\kappa_{2}T,\qquad\kappa_{2}=\frac{5}{4} \, .
\ee

In principle, a similar procedure can be carried out to calculate the $n$th moment of the distribution of $A$ for general $n=1,2,\dots$. However, the calculation is not so practical because it very quickly becomes very cumbersome as $n$ is increased. The double integral \eqref{AfromCdef} would be replaced by an $n$-dimensional integral over the $n$-time correlation function $\left\langle x\left(t_{1}\right)\cdots x\left(t_{n}\right)\right\rangle $.

An alternative approach, that works in many cases, is to use the DV large-deviation formalism. This involves calculating the scaled cumulant generating function (SCGF) $\lambda(k)$, defined as
$\left\langle e^{kA}\right\rangle \sim e^{T\lambda\left(k\right)}$.
When this calculation succeeds, $\lambda(k)$ gives the coefficients of the linear growth of all of the cumulants in time, 
\be
\label{kappanTime}
\left\langle A^{n}\right\rangle _{c}\simeq\kappa_{n}T \, ,
\ee
via its power-series expansion
\be
\label{kappanDef}
\lambda\left(k\right)=\sum_{n=1}^{\infty}\frac{\kappa_{n}}{n!}k^{n} \, .
\ee
Furthermore, a successful application of this approach implies that the LDP \eqref{LDP} holds, with a rate function that is given by the Legendre-Fenchel transform of $\lambda(k)$ \cite{Touchette2018}:
\be
I\left(a\right)=\sup_{k\in\mathbb{R}}\left[ka-\lambda\left(k\right)\right] \, .
\ee

To calculate $\lambda(k)$, one must find the largest eigenvalue of a tilted generator of the dynamics.
For a Brownian particle diffusing in a potential $V(x) = |x|$, this problem can be reformulated \cite{Touchette2018} as that of finding the ground-state energy $E = -\lambda$ of a quantum particle of unit mass in a tilted potential
\be
\label{Ukdef}
U_{k}\left(x\right)=\frac{V'\left(x\right)^{2}}{2}-\frac{V''\left(x\right)}{2}-ku\left(x\right)=\frac{1}{2}-\delta\left(x\right)-kx \, .
\ee
In other words, we must find the ground-state energy corresponding to the time-independent Schrödinger equation
\be
\label{Schrodinger}
-\frac{1}{2}\psi''\left(x\right)+\left[\frac{1}{2}-\delta\left(x\right)-kx\right]\psi\left(x\right)=E\psi\left(x\right) \, .
\ee
One immediately notices that, except for the particular case $k=0$, the potential $U_k(x)$ is not confining, i.e., it does not have a ground state because its spectrum is unbounded from below. Based on previous studies in other systems, including the very closely related Brownian motion with drift \cite{BH20}, but also, e.g., the Ornstein-Uhlenbeck process \cite{NT18, Smith22OU, NT22} and stochastically resetting fractional Brownian motion \cite{DH2019, SmithMajumdar22}, we expect this to cause the DV formalism to break down, and with it, also the LDP \eqref{LDP}.

Nevertheless, since a ground state does exist at $k=0$, one can calculate the energy of a quasi ground state perturbatively at $k\ll 1$. 
 Technically, this procedure is somewhat similar to calculations that are sometimes performed in quantum theory: One example is the Stark effect, where one calculates energy ``levels'' of the hydrogen atom that are perturbed  due a small external electric field, when strictly speaking there are no bound states for the perturbed system.
One way to interpret such a calculation is by regularizing the perturbed Hamiltonian so that it does have a bound state for nonzero $k$, for instance by adding reflecting walls at $|x| = L$, and then taking the limit $L \to \infty$.
In any case, the result of this perturbative treatment of the DV calculation is expected to yield a perturbative expansion of the form \eqref{kappanDef}, from which the coefficients $\kappa_n$ can be read off, predicting a linear growth of the cumulants in time (at long times).

Let us carry out this perturbative calculation explicitly for our system. At $k=0$, the ground-state energy of Eq.~\eqref{Schrodinger} is $E=0$, and the corresponding wave function is
\be
\psi\left(x\right)=e^{-\left|x\right|} \, .
\ee
Let us now treat the case $0 < k \ll 1$ (it is sufficient to consider the case of $k$ positive because of the mirror symmetry of the problem).
The Schrödinger equation \eqref{Schrodinger} is easily solved at positive and negative $x$'s separately in terms of Airy functions. Keeping in mind, e.g., the regularization scheme given above (reflecting walls at $|x| = L$), we look for a solution to Eq.~\eqref{Schrodinger} that, locally around the origin $x=0$, decays in both directions. Thus we find that
\be
\label{psisol}
\psi\left(x\right)=\begin{cases}
C_1 \text{Ai}\left(\frac{-2E-2kx+1}{2^{2/3}k^{2/3}}\right), & x<0,\\[2mm]
C_2 \text{Bi}\left(\frac{-2E-2kx+1}{2^{2/3}k^{2/3}}\right), & x>0,
\end{cases}
\ee
where $\text{Ai}(z)$ and $\text{Bi}(z)$ are the Airy functions of the first and second kind respectively, and $C_1$ and $C_2$ are constants. 
Requiring continuity of the wave function $\psi(x)$ at $x=0$ and that the wave function's first derivative jumps at $x=0$ according to
\be
-\frac{1}{2}\left[\psi'\left(0^{+}\right)-\psi'\left(0^{-}\right)\right]-\psi\left(0\right)=0 \, ,
\ee
we obtain
$E$ as a function of $k$ via the solution to the transcendental equation (see Appendix \ref{app:DV} for the details)
\be
\label{Eofk2}
\frac{k^{1/3}}{2^{2/3}\pi}=\text{Ai}\left(\frac{1-2E}{2^{2/3}k^{2/3}}\right)\text{Bi}\left(\frac{1-2E}{2^{2/3}k^{2/3}}\right) \, .
\ee

\begin{figure}[ht]
\includegraphics[width=0.51\linewidth,clip=]{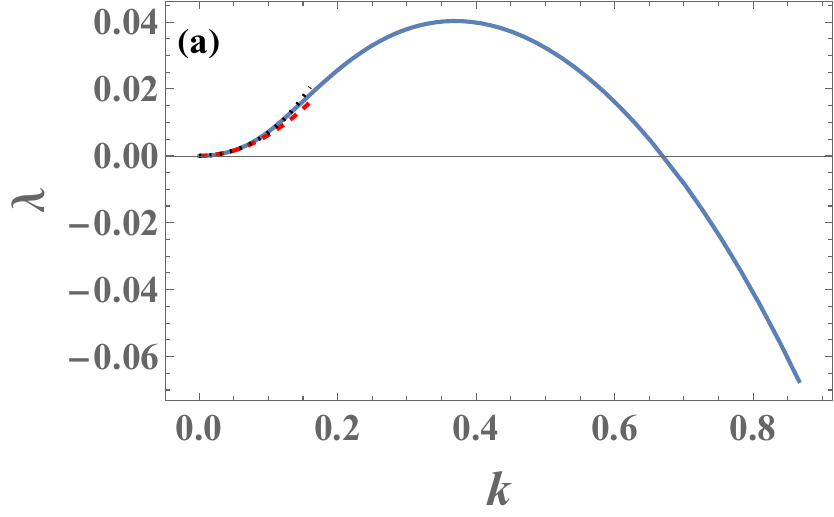}
\hspace{2mm}
\includegraphics[width=0.46\linewidth,clip=]{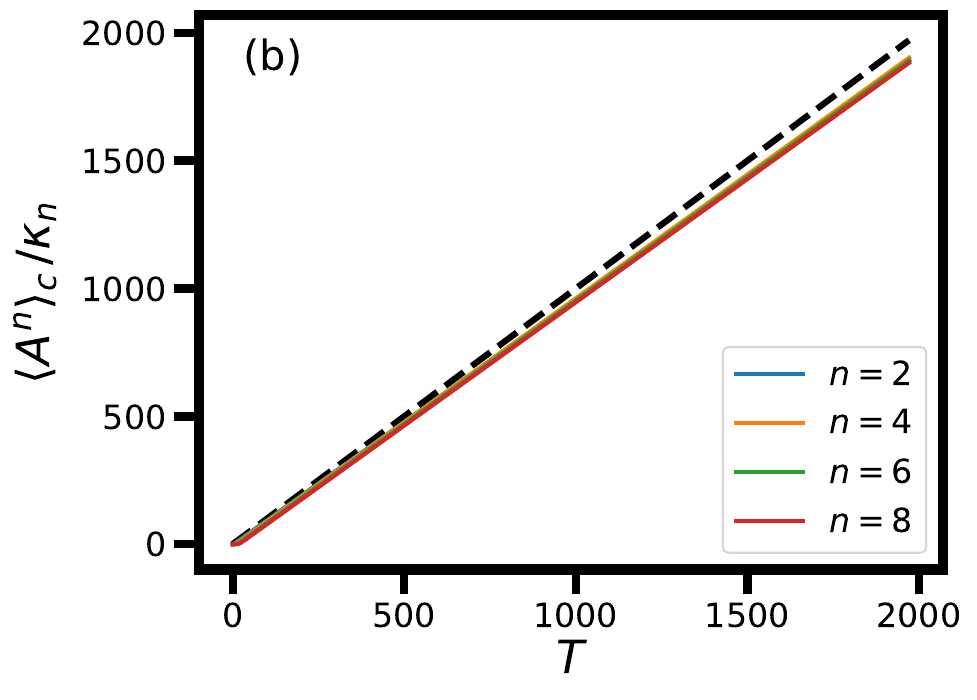}
\caption{(a) Solid line: $\lambda = -E$ as a function of $k$, as given by the numerical solution to Eq.~\eqref{Eofk2}. The dashed line corresponds to the quadratic approximation $\lambda \sim k^2$ around $k=0$, and the dotted line represents the approximation obtained by adding the $\sim k^4$ term.
(b) Solid lines: The ratio between the four lowest nonvanishing cumulants and their coefficients $\kappa_n$, which we obtained numerically \cite{Smith24Numerical}. The curves appear to agree very well with the prediction $\left\langle A^{n}\right\rangle _{c}\simeq\kappa_{n}T$ (dashed line).}
\label{fig:cumulants}
\end{figure}

A numerical solution to Eq.~\eqref{Eofk2} is plotted in Fig.~\ref{fig:cumulants}. However, Eq.~\eqref{Eofk2} is only expected to give a physically-meaningful result in the limit $k \to 0$, i.e., when $k$ is treated as a perturbative parameter. It is convenient to write its solution in a parametric form, as follows. Denoting the argument of the Airy functions in \eqref{Eofk2} by
$\eta=\left(1-2E\right)/\left(2k\right)^{2/3}$,
we have
\bea
\label{kOfEta}
k\left(\eta\right)&=&4\left[\pi\text{Ai}\left(\eta\right)\text{Bi}\left(\eta\right)\right]^{3}\, , \\[1mm]
\label{EOfEta}
E\left(\eta\right)&=&\frac{1-4\eta\left[\pi\text{Ai}\left(\eta\right)\text{Bi}\left(\eta\right)\right]^{2}}{2} \, .
\eea
From the definition of $\eta$, it is clear that the limit $k \to 0$ corresponds to $\eta \to \infty$ (to remind the reader $E(k=0)=0$). Thus, the $k \to 0$ limiting behavior of $\lambda = -E$ can be deduced by taking the asymptotic behaviors of Eqs.~\eqref{kOfEta} and \eqref{EOfEta} at $\eta \to \infty$, which read
\bea
k\left(\eta\right)&=&\frac{1}{2\eta^{3/2}}+\frac{15}{64\eta^{9/2}}+\frac{3615}{4096\eta^{15/2}}+\frac{1311175}{131072\eta^{21/2}}+\dots,\\[1mm]
\lambda\left(\eta\right)&=&-E\left(\eta\right)=\frac{5}{32\eta^{3}}+\frac{295}{512\eta^{6}}+\frac{13475}{2048\eta^{9}}+\frac{20487775}{131072\eta^{12}} + \dots .
\eea
These two equations then yield the following power-series expansion of $\lambda$ as a function of $k$:
\be
\label{lambdakSol}
\lambda\left(k\right)=-E\left(k\right)=\frac{5}{8}k^{2}+\frac{55}{8}k^{4}+\frac{10625}{32}k^{6}+\frac{1078125}{32}k^{8}+ \dots.
\ee
Comparing this power series expansion to Eq.~\eqref{kappanDef}, we find
\be
\label{kappasols}
\kappa_{2}=\frac{5}{8}2!=\frac{5}{4},\qquad\kappa_{4}=\frac{55}{8}4!=165,\qquad\kappa_{6}=\frac{10625}{32}6!=\frac{478125}{2}, \kappa_{8}=\frac{1078125}{32}8!=1358437500 \, ,
\ee
describing the linear growth \eqref{kappanTime} of the four lowest nonzero cumulants with time.
We tested the predictions \eqref{kappasols} by comparing them to numerical computations of the cumulants, with very good agreement, see Fig.~\ref{fig:cumulants} (b) (we attribute the small discrepancies to numerical error).
The numerical results are based on a numerical computation of $P(A;T)$ by solving of the joint Fokker-Planck equation for $x$ and $A$, whose details will be given separately \cite{Smith24Numerical}. We use the numerically-computed $P(A;T)$ to compute the moments of the distribution, and from them, the cumulants (e.g., the fourth cumulant is given by 
$\left\langle A^{4}\right\rangle _{c}=\left\langle A^{4}\right\rangle -3\left\langle A^{2}\right\rangle ^{2}$ where we used that all of the odd moments vanish due to mirror symmetry).

A few remarks are in order. First of all, the coefficients in the power series \eqref{lambdakSol} appear to grow very rapidly, signaling that its radius of convergence is probably zero. 
This is despite the fact that the solution to Eq.~\eqref{Eofk2} exists at all values of $k>0$. However, as explained above, $\lambda(k)$ is only expected to be physically meaningful in the limit $k \to 0$ (where it reduces to its power-series expansion).
%
It is worth mentioning that a very similar situation, in which the LDP \eqref{LDP} is violated but nevertheless the cumulants grow linearly in time (at long times) was also observed in \cite{SmithMajumdar22} when considering the distribution of the area under a stochastically-resetting Brownian motion.

The cumulants give a very accurate description of the distribution of $A$ in the typical-fluctuations regime at $T \gg 1$. The linear growth of all cumulants in time implies that typical fluctuations of $A$ are described by a Gaussian distribution with zero mean, and a variance which is given by $\kappa_2 T$, i.e.,
\be
\label{Gaussian}
P\left(A;T\right)\simeq\frac{1}{\sqrt{2\pi\kappa_{2}T}}e^{-A^{2}/\left(2\kappa_{2}T\right)} \, .
\ee
The higher cumulants give the (relatively small) corrections to this Gaussian behavior. However, in contrast to the ``usual" situation, $\lambda(k)$ does not describe large deviations of $A$ through a Legendre transform. A  detailed analysis of the large-deviation regime, which we perform next, requires one to combine the (perturbative) DV theory with other large-deviation techniques.

\section{Very large deviations $A \sim T^2$}
\label{sec:veryLarge}

Very large fluctuations of $A$ are described well by the optimal fluctuation method (OFM), also known as the instanton method or weak-noise approximation. This involves approximating $P(A;T)$ by the probability of observing the most probable realization of the process $x(t)$ conditioned on a given $A$. In the context of a particle undergoing Brownian motion, this theory takes a form that is very similar to geometrical optics \cite{Meerson19, AiryDistribution20, SmithMeerson2019, MeersonSmith2019, Meerson23RA, BarMeerson2023}.

The result in this regime has been obtained previously in the Mathematics literature \cite{DM10} in a closely-related system by considering the continuum limit of a discrete-time random walk, so we will confine ourselves here to a brief, physical derivation whose goal is to keep the paper self contained.
Our starting point is to write the probability of a realization of the white noise $\xi(t)$ (up to a normalization factor),
$\mathbb{P}\left[\xi\left(t\right)\right] \sim e^{-\frac{1}{2}\int_{0}^{T}\xi^{2}\left(t\right)dt}$.
Eliminating $\xi$ from Eq.~\eqref{LangevinOrig}, we can similarly write the probability of a realization of $x(t)$, in the physical variables, as
\be
\label{SDimensionalDef}
\mathbb{P}\left[x\left(t\right)\right]\sim e^{-S\left[x\left(t\right)\right]},\qquad S\left[x\left(t\right)\right]=\frac{1}{4D}\int_{0}^{T}\left[\dot{x}\left(t\right)+\mu \, \text{sgn}\left(x\right)\right]^{2}dt \, .
\ee
In the weak-noise limit $D \to 0$ (which is the limit on which we will focus on for the remainder of this section), $P(A;T)$ is dominated by the realization $x(t)$ that minimizes the ``action'' $S$, constrained on the area \eqref{AreaDef} attaining the value $A$.

Analyzing this problem now in the dimensionless variables, we find from Eq.~\eqref{RescalingPAT} that the limit $D \to 0$ is in fact equivalent to the joint limit $A,T \to \infty$ with fixed $A/T^2$, i.e., the OFM becomes asymptotically exact at $A \sim T^2$ in the long-time limit. 
Keeping this in mind, we now switch to the rescaled variables, in which Eq.~\eqref{SDimensionalDef} becomes
\be
\label{sDef}
\mathbb{P}\left[x\left(t\right)\right]\sim e^{-s\left[x\left(t\right)\right]},\qquad s\left[x\left(t\right)\right]=\frac{1}{2}\int_{0}^{T}\left[\dot{x}\left(t\right)+\text{sgn}\left(x\right)\right]^{2}dt \, .
\ee
To find $P(A;T)$, one must minimize $s[x(t)]$ under the initial condition $x(0)=0$ and under the constraint $\int_{0}^{T}x\left(t\right)dt=A$. The latter constraint is easily incorporated via a Lagrange multiplier, i.e., by minimizing the modified action
\be
\label{sNuDef}
s_{\nu}\left[x\left(t\right)\right]=\int_{0}^{T}\left\{ \frac{1}{2}\left[\dot{x}+\text{sgn}\left(x\right)\right]^{2}+\nu x\right\} dt \, ,
\ee
where $\nu$ is a Lagrange multiplier whose value is ultimately set by the constraint $\int_{0}^{T}x\left(t\right)dt=A$.
Since $x(t=T)$ is not specified, there is a ``free" boundary condition at $t=T$
\be
\label{BCEnd}
\left.\frac{\partial L}{\partial\dot{x}}\right|_{t=T}=\left[\dot{x}+\text{sgn}\left(x\right)\right]_{t=T}=0
\ee
where $L$ is the Lagrangian corresponding to $s_{\nu}$.

\begin{figure}[ht]
\includegraphics[width=0.485\linewidth,clip=]{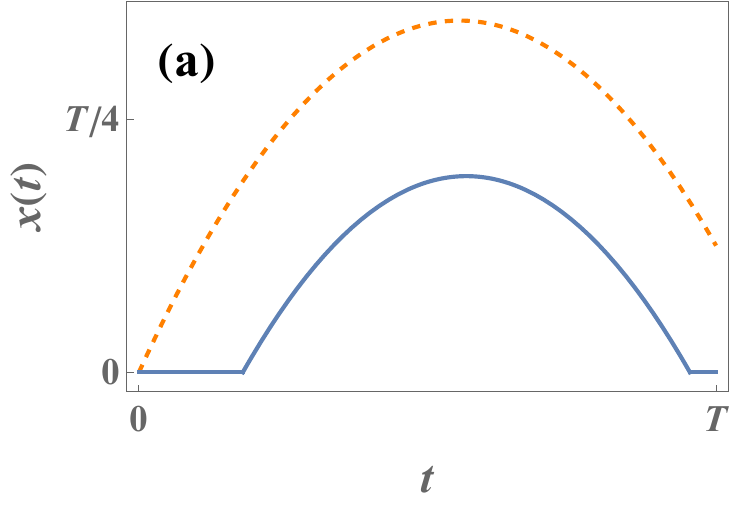}
\hspace{1mm}
\includegraphics[width=0.465\linewidth,clip=]{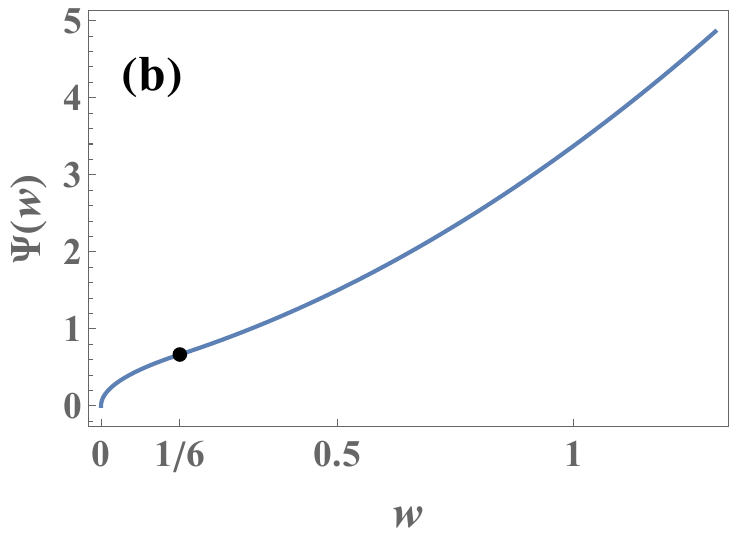}
\caption{(a) Optimal trajectories that give the leading-order contribution to $P(A;T)$ in the regime of very large deviations, $A\sim T^2$. Trajectories are plotted for a subcritical value  $A / T^2 = 4$ and supercritical value  $A/T^2 = 10$ (solid and dashed lines respectively).
(b) The large-deviation function $\Psi(w)$ that describes $P(A;T)$ in this regime, see Eqs.~\eqref{PsiScaling} and \eqref{PsiSol}. At the critical value $w=1/6$ (marked by the dot in the figure), $\Psi(w)$ exhibits a singularity: Its third derivative jumps. We interpret this as a third-order dynamical phase transition.}
\label{fig:GeometricalOpticsAndPsi}
\end{figure}

We now solve this minimization problem. Due to the mirror symmetry of the problem, let us assume for simplicity that $A>0$.
There are two classes of minimizers. The first class is trajectories $x(t)$ for which $x(t) > 0$ for all $t>0$. In this case, the boundary condition \eqref{BCEnd} becomes $\dot{x}\left(t=T\right)=-1$.
Then the Euler-Lagrange equation (for $t>0$) is simply $\ddot{x}=\nu$, and its solution under the boundary conditions $x(0)=0$ and $\dot{x}\left(t=T\right)=-1$ is
$x\left(t\right)=\frac{\nu t^{2}}{2}-\left(\nu T+1\right)t$.
We now find the value of $\nu$ via
\be
A=\int_{0}^{T}x\left(t\right)dt=-\frac{T^{2}\left(2\nu T+3\right)}{6}
\ee
and thus
\be
\label{xOftSuper}
x\left(t\right)=\left[\frac{3\left(2A+T^{2}\right)}{2T^{2}}-1\right]t-\frac{3\left(2A+T^{2}\right)}{4T^{3}}t^{2} \, ,
\ee
see Fig.~\ref{fig:GeometricalOpticsAndPsi}(a).
This solution is only correct if it indeed satisfies $x(t)>0$ for all $T>t>0$, and this in turn implies that $A \ge T^{2}\!/6$.
Plugging Eq.~\eqref{xOftSuper} into \eqref{sDef}, one finds that the action of this solution is
\be
\label{sSuper}
s=\frac{3\left(2A+T^{2}\right)^{2}}{8T^{3}} \, .
\ee

The second class of minimizer  is of a slightly different nature, and it corresponds to the case $0 < A < T^{2}/6$. To find it, it is useful to notice that the Lagrangian in \eqref{sNuDef} does not explicitly depend on $t$, and therefore the Jacobi energy function is conserved:
\be
\label{EDef}
E=\dot{x}\frac{\partial L}{\partial\dot{x}}-L=\frac{1}{2}\left[\dot{x}-\text{sgn}\left(x\right)\right]\left[\dot{x}+\text{sgn}\left(x\right)\right]-\nu x=\text{const} \, .
\ee
For $0 < A < T^{2}/6$, the minimizer is of zero ``energy'', i.e., $E=0$, and it is composed of segments of two different types: (i) segments in which $x(t)\equiv0$ [recall that the sign function is defined such that $\text{sgn}\left(0\right)=0$], (ii) segments in which $x(t)>0$. The minimizer is of course a continuous function of $t$, so the two types of segments can alternate.
A segment of type (ii), lasting for the time interval $[t_1, t_2]$, is a parabolic ``bump''
$x\left(t\right)=-\nu\left(t-t_{1}\right)\left(t_{2}-t\right)/2$.
Within segments of type (ii), one has
$E=\frac{1}{2}\left(\dot{x}^{2}-1\right)-\nu x=0$ and thus, it must satisfy $\dot{x}\left(t_{1}\right)=1$ and $\dot{x}\left(t_{2}\right)=-1$. It then follows 
that the duration of the ``bump'' must be
$t_{2}-t_{1}=-2/\nu$.
Moreover, one finds that the area under the bump is
\be
A_{12}=\int_{t_{1}}^{t_{2}}x\left(t\right)dt=-\frac{\nu\left(t_{2}-t_{1}\right)^{3}}{12}=\frac{2}{3\nu^{2}}
\ee

The simplest such solution is a solution with a single bump. In this case one has $A_{12}=A$, so the trajectory takes the form
\be
x\left(t\right)=\begin{cases}
\frac{1}{\sqrt{6A}}\,\left(t-t_{1}\right)\left(t_{2}-t\right), & t\in\left[t_{1},t_{2}\right],\\[2mm]
0, & t\in\left[0,t_{1}\right]\cup\left[t_{2},T\right],
\end{cases}
\ee
where $t_{2}-t_{1}=\sqrt{6A}$, see Fig.~\ref{fig:GeometricalOpticsAndPsi}(a).
For this solution, the only nonzero contribution to the action comes from the interval $[t_1, t_2]$, and one finds
\be
\label{sSub}
s=\frac{1}{2}\int_{t_{1}}^{t_{2}}\left[\dot{x}\left(t\right)+1\right]^{2}dt=\frac{2^{3/2}\sqrt{A}}{3^{1/2}} \, ,
\ee
which is independent of $T$ and of $t_1$. 
A very similar analysis shows that in general, the action for a solution with $n$ bumps is $s=\frac{2^{3/2}\sqrt{nA}}{3^{1/2}}$, and thus the solution with a single bump ($n=1$) is optimal: It minimizes the action $s$ under the constraints.
Note that there is a degeneracy of solutions, corresponding to different choices of $t_1$. All of these solutions have the same action. However, this degeneracy has no effect on the leading order calculation of $P(A;T)$ that we are performing here.

Putting Eqs.~\eqref{sSuper} and \eqref{sSub} together, we obtain the result given above in Eqs.~\eqref{PsiScaling} and \eqref{PsiSol}.
The qualitatively different behaviors of the optimal path in the subcritical ($A<T^2/6$) and supercritical ($A \ge T^2/6$) regimes lead to a singularity in the large-deviation function $\Psi(w)$ at the corresponding critical value $w=1/6$: The third derivative $\Psi'''(w)$ jumps at the critical point. $\Psi(w)$ is plotted in Fig.~\ref{fig:GeometricalOpticsAndPsi}(b).

Finally, note that taking the limit $A \ll T^2$ in this regime, the ``bump" becomes localized in time (in the sense that its duration $\sqrt{6A}$ is much shorter than $T$). This is somewhat reminiscent of the ``big-jump principle", a well-known phenomenon in the context of large deviations of sums of independent, identically distributed (i.i.d.) random variables, that occurs if the PDF of each of the summands decays slower than an exponential \cite{Chistyakov64,Foss13,Denisov08,Geluk09,Clusel06, BCV10,BUV14,VBB19, WVBB19,Gradenigo13,Barkai20, MKB98, BBBJ2000, EH05, MEZ05, EMZ06, Majumdar10, CC12, ZCG, Corberi15, BVB24}).

\section{Moderately large deviations $A \sim T^{2/3}$}
\label{sec:moderate}

%
As shown above, in the typical-fluctuations regime [denoted below as regime (i)], $P(A;T)$ scales as $-\ln P(A;T) \sim A^2 / T$, while in the subcritical part of the very-large-deviations regime [denoted below as regime (ii)], $-\ln P(A;T) \sim \sqrt{A}$. It is therefore natural to expect some intermediate behavior at values of $A$ for which these two predictions are of the same order of magnitude, i.e., $A \sim T^{2/3}$. This is indeed the case, and we refer to this regime as the moderately-large-deviation regime [and denote it regime (iii)].

The analyses given above of the two regimes (i) 
and (ii) 
are extremely helpful to understand the regime (iii), 
as the description of regime (iii) combines elements from regimes (i) and (ii). It is particularly important to understand the scenarios that dominate the contribution to $P(A;T)$ in the different regimes. This is closely related to understanding the conditioned process --- the process $x(t)$ conditioned on a given value of $A$.

In regime (i), the standard Donsker-Varadhan picture is valid: The conditioned process $x(t)$ is stationary, i.e., its statistical properties are homogeneous in time, and the realizations that dominate the contribution to $P(A;T)$ are such that the integral $A=\int_0^Tx(t)dt$ grows (over long timescales) at a constant rate in time.
In contrast, if one takes the limit $A \ll T^2$ in the result of regime (ii), the dominant realizations $x(t)$ are extremely inhomogeneous in time: The dominant contribution to $A$ comes from a localized ``bump" around some intermediate time $t_1 \in (0,T)$ (see above).

In regime  (iii), one may consider a combined scenario, in which there are two contributions to $A$: one that is temporally localized, and another that is temporally homogeneous. Let us denote them by $A_1$ and  $A_0$, respectively such that $A=A_1 + A_0$, and we will assume for now that they are of the same order of magnitude, i.e., $A_0 \sim A_1 \sim A$, and in addition, we assume the scaling $A \sim T^{2/3}$ as explained below.
The mechanisms behind the two contributions operate on very different temporal and spatial scales: 
$A_0$ is attained over a period of time $\simeq \! T$, during which the particle's mean position is $\simeq \! A_0 / T \sim T^{-1/3}$, whereas $A_1$ is attained during a period $ \sim \! \sqrt{A_1} \sim T^{1/3}$ in which the particle's position is also of order $\sim \! \sqrt{A_1} \sim T^{1/3}$. As a result, the mechanisms that create the two contributions $A_0$ and $A_1$, in the leading order (at $T \gg 1$), do not affect each other. Thus, the
 two contributions are, in the leading order, statistically independent, and their probabilities may be calculated using the perturbative DV and geometrical optics methods, respectively, see above.
To summarize, the probability for such a combined scenario is given by%
\be
\label{pA1A}
p\left(A_0,A_{1};T\right)\sim\exp\left(-\frac{2^{3/2}\sqrt{A_1}}{3^{1/2}}-\frac{2A_{0}^{2}}{5T}\right)\,.
\ee

\begin{figure*}[ht]
\includegraphics[width=0.65\linewidth,clip=]{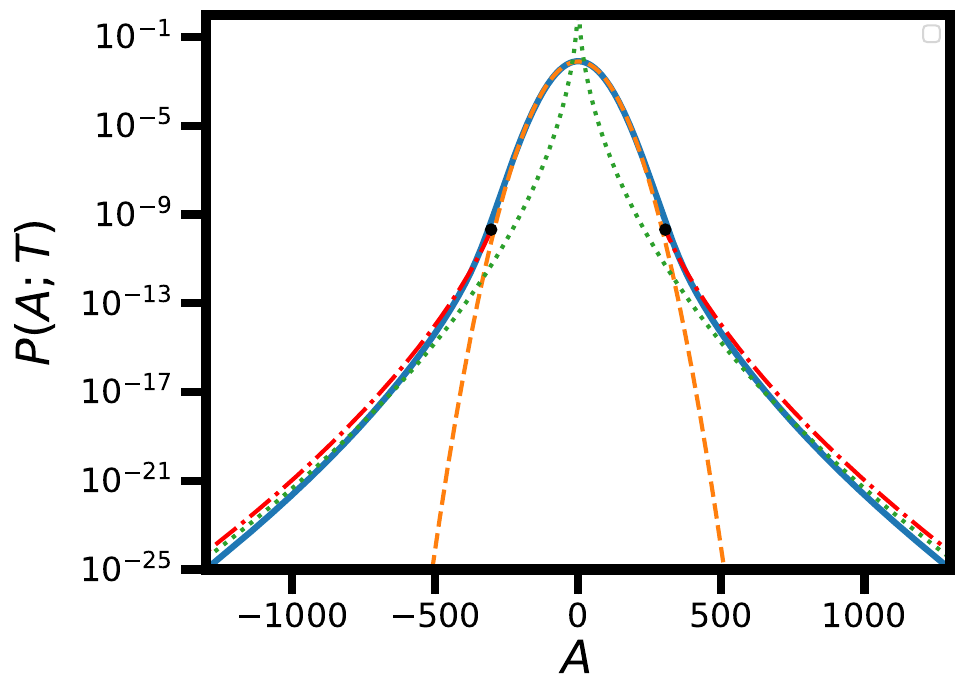}
\hspace{1mm}
\caption{$P(A;T)$ as a function of $A$ at $T=1975$, in a semi-log scaling. 
Solid line: Numerical computation (whose details will be presented separately \cite{Smith24Numerical}). 
Dashed line: Gaussian asymptotic behavior \eqref{Gaussian} that describes the typical-fluctuations regime. 
Dotted line: The asymptotic behavior $P\left(A;T\right)\sim e^{-2^{3/2}\sqrt{A/3}}$ [see Eq.~\eqref{sSub}] that describes the subcritical part of the very-large-deviations regime,  $|A| / T^2 < 6$. 
Dot-dashed line: The prediction \eqref{fScaling} and \eqref{fsol} for the supercritical part of the moderately-large-deviations regime, $|A| / T^{2/3} > y_c$ (in the subcritical part, this prediction coincides with the dashed line). The critical points  $|A| / T^{2/3} =y_c$ corresponding to $A \simeq \pm 380$ are marked by the fat dots.
}
\label{fig:PA}
\end{figure*}

One may now evaluate the PDF $P(A;T)$ of $A=A_0+A_1$ by using that it is (approximately) the convolution of the calculating the PDFs of $A_0$ and $A_1$, i.e.,
\be
\label{PATIntegralA1}
P(A;T) \sim \int_0^A p\left(A-A_1,A_{1};T\right) dA_1 \, ,
\ee
where the (finite) integration limits are introduced for convenience (clearly the dominant contribution to the integral will come from $A_0$ and $A_1$ that are both nonnegative).
We now rescale $A=T^{2/3}y$, and thus Eq.~\eqref{PATIntegralA1} becomes, in the leading order (after changing the integration variable $A_1 = T^{2/3} z$),
\be
\label{PATIntegralz}
P(A;T)\sim\int_{0}^{y}\exp\left[-T^{1/3}\left(\frac{2^{3/2}\sqrt{z}}{3^{1/2}}+\frac{2\left(y-z\right)^{2}}{5}\right)\right]dz\,.
\ee
Exploiting the large parameter $T^{1/3} \gg 1$, we apply the saddle-point approximation in Eq.~\eqref{PATIntegralz}, yielding the result reported above for the moderately-large-deviations regime, see Eqs.~\eqref{fScaling} and \eqref{fsol}.
This result exhibits good agreement with a numerical computation \cite{Smith24Numerical} of $P(A;T)$ at $T=1975$, see Fig.~\ref{fig:PA}.

Remarkably, the rate function $f(y)$ exhibits a corner singularity at the critical value $y_{c}={15}^{2/3}/2^{4/3}=2.41372\dots$. In the subcritical regime $y<y_c$, the minimizer in Eq.~\eqref{fsol} is $z=0$, corresponding to $A_1=0$ in the analysis above. This is the (temporally) homogeneous phase.
In the supercritical regime $y>y_c$, the minimizer in Eq.~\eqref{fsol} is at a nonzero value of $z$, and this is thus the condensed phase. At $y \gg 1$ (corresponding to $A \gg T^{2/3}$), the minimizer is at $z\simeq y$ so one simply has
$f\left(y\right)\simeq2^{3/2}\sqrt{y/3}$, matching smoothly with the parabolic behavior of $\Psi(w)$ at $w \to 0$. For completeness, we present a brief derivation of these properties of $f(y)$ in Appendix \ref{app:fofy}.

 Finally, let us briefly outline an alternative derivation of the moderate-large deviations regime \cite{Smith22OU}. For any $N=1,2,\dots,$ One can write the observable $A$ in the form
$A=\sum_{i=1}^{N}\mathcal{A}_{i}$
where
$\mathcal{A}_{i}=\int_{\left(i-1\right)T/N}^{iT/N}x\left(t\right)dt$.
Choosing $N$ such that $N \ll T$, the random variables $\mathcal{A}_1, \dots, \mathcal{A}_N$ become approximately i.i.d., since the duration $T/N$ of each of the integrals that defines the $\mathcal{A}_{i}$'s is much longer than the correlation time of the system (which is of order unity).
Requiring, in addition, that $N \gg 1$, one can now analyze the problem using standard tools that describe large deviations of sums of a large number of i.i.d. random variables.

To apply these tools, it is sufficient to calculate the variance and the near-tail behavior of each of the $\mathcal{A}_{i}$'s. This is easily done by using the same analysis as we did when we performed such a calculation for $A$ itself in the previous sections; The result is
$\text{Var}\mathcal{A}_{i}\simeq5T/4N$
and the near tail behaves as
$P\left(\mathcal{A}_{i}\right)\sim e^{-2^{3/2}\sqrt{\mathcal{A}_{i}}/\sqrt{3}}$.
Plugging these behaviors into the general formulas found e.g., in Ref. \cite{BKLP20} leads precisely to the same results \eqref{fScaling} and \eqref{fsol}.

\section{Absolute area}
\label{sec:absoluteArea}

Let us briefly consider fluctuations of the absolute area
\be
\label{Bdef}
B=\int_{0}^{T}\left|x\left(t\right)\right|dt\,,
\ee
corresponding to $u(x)=|x|$ in Eq.~\eqref{DynamicalObservable}. 
The absolute area \eqref{Bdef} is (exactly) equivalent to the area swept by a Brownian particle with constant drift toward a reflecting wall at the origin. This problem was studied in Ref.~\cite{BH20}.
$B$ behaves a little differently to the area $A$ studied above.
To begin with, it has a nonzero mean value \cite{BH20}
\be
\bar{B}\equiv\left\langle B\right\rangle =\frac{T}{2}
\ee
and its distribution $P(B;T)$ is highly asymmetric around $\bar{B}$.
Furthermore, Eq.~\eqref{Ukdef} for the effective DV potential now gives way to
\be
\label{UkofBdef}
U_{k}\left(x\right)=\frac{1}{2}-\delta\left(x\right)-k\left|x\right| \, ,
\ee
which is a confining potential for all $k\le 0$.
As a result, $\lambda(k)$ can be calculated using the standard (non-perturbative) DV theory at $k\le0$, and from it one finds that the left tail of the distribution, $B < \bar{B}$, is described by the standard LDP
$P\left(B;T\right)\sim e^{-T\tilde{I}\left(B/T\right)}$.
where $\tilde{I}(b)$ is nontrivial for $0<b<1/2$ (describing the left tail of the distribution), and vanishes at $b=1/2$ \cite{BH20}.
In Appendix \ref{app:absoluteArea}, we use the transcendental equation obtained in \cite{BH20} for $\lambda(k)$ to obtain $\tilde{I}(b)$ (at $0<b<1/2$) in a parametric form, and show that its asymptotic behavior at $1/2 - b \ll 1$ is
\be
\tilde{I}\left(b\right)\simeq\left(b-\frac{1}{2}\right)^{2} \, .
\ee
As a result, typical fluctuations follow a Gaussian distribution with a variance that grows linearly in time as $\text{Var}\left(B\right)\simeq T/2$.
In analogy with our results given above for $P(A;T)$, all the cumulants of $B$ grow linearly in time with coefficients that can be obtained analytically from the expansion of $\lambda(k)$ at $k\to0^-$, see Appendix \ref{app:absoluteArea} for the first few coefficients.

We now turn to the investigation of large deviations $B > \bar{B}$, following similar steps as in the analysis of $P(A;T)$.
The very-large-deviations regime $B\sim T^2$ is described by geometrical optics, and one finds that in this regime $P(B;T)$ and $P(A;T)$ coincide in the leading order, i.e., $P(B;T)$ is also given by Eqs.~\eqref{PsiScaling} and \eqref{PsiSol}. The reason for this is that the optimal paths $x(t)$ calculated in section \ref{sec:veryLarge} satisfy $x(t) \ge 0$ and are therefore also optimal for the case in which the absolute area is considered (additional optimal paths may be obtained by flipping the sign of $x$, but they have the exact same action and therefore do not affect the result in the leading order).

Next, we consider the regime of moderately-large fluctuations $B > \bar{B}$. Following very similar arguments to those given in section \ref{sec:moderate}, we find that this regime is given by
$B-\bar{B}\sim T^{2/3}$. Within this regime, $P(B;T)$ can be obtained using a very similar analysis to that given in Eqs.~\eqref{pA1A}-\eqref{PATIntegralz} (we do not go into the details here). The result is again an anomalous scaling,
\be
P\left(B;T\right)\sim\exp\left[-T^{1/3}\tilde{f}\left(\frac{B-T/2}{T^{2/3}}\right)\right] \, ,
\ee
with a rate function 
\be
\label{ftildeSol}
\tilde{f}\left(y\right)=\min_{0\le z\le y}\left[\frac{2^{3/2}\sqrt{z}}{3^{1/2}}+\left(y-z\right)^{2}\right]\,.
\ee
The only difference between $f$ and $\tilde{f}$ is due to the numerical coefficients in front of the terms $(y-z)^2$ in Eqs.~\eqref{fsol} and \eqref{ftildeSol} respectively. This, in turn, is a result of the difference between the numerical coefficients that describe the linear growth of the variances of $A$ and $B$ in time.

\section{Higher dimension}
\label{sec:higherDim}

Let us now discuss the extension of our results to higher spatial dimension. We consider a Brownian particle $\vect{r}(t)$ in $d$ dimensions (in rescaled space and time) trapped by a potential $V(\vect{r}) = r$ where $r=|\vect{r}|$,
\be
\label{LangevinHighd}
\dot{\vect{r}}\left(t\right)=-\vect{r}\left(t\right)/r+\vect{\xi}\left(t\right) \, .
\ee
Now $\vect{\xi}(t)$ is a $d$-dimensional white noise with
$\left\langle \ensuremath{\vect{\xi}}\left(t\right)\right\rangle =0$
and
$\left\langle \xi_{i}\left(t\right)\xi_{j}\left(t'\right)\right\rangle =\delta_{ij}\delta\left(t-t'\right)$.
%
%
Let $x(t)$ denote the projection of the particle's position $\vect{r}(t)$ on some specified direction in space (the $x$ direction).
We consider fluctuations of the area under the $x$-component,
\be
\mathcal{A}=\int_{0}^{T}x\left(t\right)dt \, .
\ee

Let us follow a similar analysis to the one that we applied in the case $d=1$.
The (standard) DV formalism requires one to calculate the ground-state energy of a quantum particle in $d$ dimensions in the effective (tilted) potential%
\footnote{For $d=1$, one must add to the right-hand side of Eq.~\eqref{UkHighDim} the term $-\delta(x)$, as we did in Eq.~\eqref{Ukdef}.}
 \cite{Touchette2018}
\be
\label{UkHighDim}
U_{k}\left(\vect{r}\right)=\frac{\left[\nabla V\left(\vect{r}\right)\right]^{2}}{2}-\frac{\nabla^{2}V\left(\vect{r}\right)}{2}-kx=\frac{1}{2}-\frac{d-1}{2r}-kx \, .
\ee
It is clearly seen that
$U_{k=0}\left(\vect{r}\right)$
may be interpreted as the Coulomb potential, corresponding to a ($d$-dimensional) hydrogen atom, while if $k\ne0$, the term $kx$ may be interpreted as the effect of an additional constant external electrostatic field (in the $x$ direction).
Again, one finds that the tilted potential is not confining, but one can obtain a perturbative expansion of the ground state energy with respect to the small parameter $k$. This is the generalization of the Stark effect to $d$ dimensions. The first-order term vanishes due to mirror symmetry, while the second order yields the coefficient $\kappa_2$ in the expansion $\lambda(k) = \kappa_2 k^2 /2! + \dots$.
In dimension $d=3$, for instance, the result is well known \cite{Bethe57}, and is given by
$\lambda = -E=9k^{2}/4$, leading to $\kappa_2 = 9/2$.
Typical fluctuations thus follow a Gaussian distribution whose variance is given by $\simeq \kappa_2 T$. Higher-order corrections to this behavior can, in principle, be obtained by going to higher orders in perturbation theory, as done above for $d=1$.

There is again a very-large-deviations regime $\mathcal{A}\sim T^2$, in which the geometrical-optics formalism may be applied. This calculation yields optimal paths for which the component $x(t)$ coincides with the optimal trajectories found above for the $d=1$ case, while the component of $\vect{r}(t)$ that is perpendicular to $x$ vanishes identically. As a result, $P(\mathcal{A};T)$ coincides, in the leading order, with $P(\mathcal{A};T)$ as it is given in Eqs.~\eqref{PsiScaling} and \eqref{PsiSol}.

Finally, one finds that there is a moderate-fluctuations regime $\mathcal{A} \sim T^{2/3}$ in which $P\left(\mathcal{A};T\right)$ is obtained again by a combination of the perturbative DV and geometrical optics methods. The result of this calculation is the anomalous scaling
\be
P\left(\mathcal{A};T\right)\sim e^{-T^{1/3}f_{d}\left(\mathcal{A}/T^{2/3}\right)}
\ee
where 
\be
f_{d}\left(y\right)=\min_{0\le z\le y}\left[\frac{2^{3/2}\sqrt{z}}{3^{1/2}}+\frac{1}{2\kappa_{2}}\left(y-z\right)^{2}\right]
\ee
(the $d$-dependence enters through $\kappa_2$).
The qualitative features of $f(y) = f_{d=1}(y)$ described above hold in $d>1$ as well.
In particular, it exhibits a first-order dynamical phase transition due to a condensation phenomenon.

\section{Summary and discussion}
\label{sec:discussion}

To summarize, we studied the full distribution $P(A;T)$ of the area $A$ swept under a Brownian motion trapped by an external potential $\sim|x|$ up to time $T$, in the long-time limit $T\to \infty$. We uncovered three regimes of interest.
In the typical-fluctuations regime, $P(A;T)$ displays standard scaling. It is described by a Gaussian distribution whose variance grows linearly with $T$. Corrections to this behavior are given by higher cumulants of the distribution, which also grow linearly with $T$.
However, large deviations of $A$ do not follow the usual large-deviations principle.
Instead, we found that there are two nontrivial large-deviations regimes, $A\sim T^{2/3}$ and $A\sim T^2$ (the latter of which had already been extensively studied previously in a closely-related setting \cite{DM10}), in which anomalous scalings are observed. We calculated the exact large-deviation functions that describe $P(A;T)$ in these two regimes and found that they both exhibit singularities, that we interpreted as dynamical phase transitions of first and third order, respectively.
We then extended our analysis by studying the absolute area and to higher dimension.

In Ref.~\cite{BBRZ22}, a generalization of the problem in which $A$ is the time-averaged $p$th moment of the distribution, it was shown that a different type of anomalous LDP holds in the regime $A \sim T$, with a rate function that is exactly given by a power law $I\left(a\right)\propto a^{1/\left(1+p\right)}$. For our case $p=1$, this result reads $I\left(a\right)\propto\sqrt{a}$, in agreement with our result for the intermediate regime $T^{2/3} \ll A \ll T^2$. However, in this system the scaling regime $A\sim T$ does not capture any of the interesting behaviors that are observed outside it (e.g., dynamical phase transitions). 

 Some of our results in the typical-fluctuations regime (where we applied our perturbative DV approach) and in the moderately-large deviations regime (where some simplifying assumptions were made) do not have a mathematically-rigorous theoretical foundation, analogous e.g. to the standard DV formalism or the geometrical optics which is based on the Martin-Siggia-Rose formalism. It would be useful to develop such rigorous frameworks and to find the precise conditions under which the tools developed here may be applied to other systems.
Furthermore, we numerically verified our theoretical predictions for $P(A;T)$ in these regimes but we were not able to numerically verify our physical interpretation in terms of the realizations of the process $x(t)$ that dominate the contribution to $P(A;T)$ for a given value of $A$, since our numerical method bypassed the need of generating individual realizations of the process. It would be useful to perform such a numerical verification by using algorithms for the efficient Monte-Carlo simulation of large deviations, such as importance sampling, see e.g. \cite{NT22}.

The fact that the cumulants grow linearly in time (at long times) despite the violation of the LDP, is rather remarkable. This feature was also observed in a system of resetting fractional Brownian motion \cite{SmithMajumdar22}. It is worth mentioning that the converse situation, in which the LDP holds, but the cumulants do not grow linearly in time, has also been observed recently \cite{Krajnik22a, Krajnik22b,Krajnik23}.
The fact that the perturbative DV theory succeeds in determining the coefficients describing the linear growth of the cumulants (despite the failure of the non-perturbative DV theory) may be a feature that is universal to a broader class of systems that display anomalous behavior similar to the one observed here \cite{Smith22OU, SmithMajumdar22}.
It is also worth noting that the rate function $f(y)$, $\tilde{f}(y)$ and $f_d(y)$ found here all  coincide, up to scaling factors, with each other and with those found in several other settings involving condensation transitions
 \cite{BKLP20, MLMS21, MGM21, Smith22OU, SmithMajumdar22}, suggesting that these transitions may all belong in the same universality class.
It would be interesting to search for similar behaviors in other systems too. For instance, in deterministic, chaotic systems, for which large deviations of dynamical observables have recently attracted  renewed interest \cite{Smith22Chaos, Monthus23Chaos, RCP23, Lippolis24}.


\bigskip

\subsection*{Acknowledgments}

I am very grateful to Hugo Touchette for useful discussions and for pointing out relevant references.
I acknowledge support from the Israel Science Foundation (ISF) through Grant No. 2651/23.

\bigskip

\appendix

\section{Calculating the variance of $A$ from the correlation function of $x(t)$}
\label{app:correlationFunction}

\renewcommand{\theequation}{A\arabic{equation}}
\setcounter{equation}{0}

In Ref.~\cite{TSJ10}, the exact propagator was calculated for the process \eqref{LangevinOrig}. In their units, corresponding to $\mu=D=1$, it is given by
\be
p\left(x,\tau\,|\,x',0\right)=\frac{e^{-\tau/4}}{2\sqrt{\pi\tau}}e^{-\left(\left|x\right|-\left|x'\right|\right)/2}e^{-\left(\left|x\right|-\left|x'\right|\right)^{2}/\left(4\tau\right)}+\frac{e^{-\left|x\right|}}{4}\left[1+\text{erf}\left(\frac{\tau-\left(\left|x\right|+\left|x'\right|\right)}{2\sqrt{\tau}}\right)\right] \, .
\ee
From the propagator, they then extracted the exact two-time correlation function \eqref{Cdef}
\be
C\left(\tau\right)=\frac{e^{-\tau/4}}{6\sqrt{\pi\tau}}\left\{ \left[\frac{\sqrt{\pi\tau}}{2}e^{\tau/4}\text{erfc}\left(\frac{\sqrt{\tau}}{2}\right)-1\right]\left(\tau^{3}+6\tau^{2}-12\tau+24\right)+2\tau^{2}+24\right\}  \, .
\ee
Plugging this into the integral that appears in Eq.~\eqref{AfromCWithIntegral}, we find
\be
\int_{0}^{\infty}C\left(\tau\right)d\tau=5 \, .
\ee
Therefore, for $\mu=D=1$ the variance grows in time as 
\be
\label{varVsTwithD1}
\left\langle A^{2}\right\rangle \simeq10T \, .
\ee
In order to translate this result into the units that we used in the rest of the present work, we find, from the scaling form \eqref{RescalingPAT} that, in the original (dimensional) variables, the variance grows linearly in time as
\be
\label{varVsTgeneralmuD}
\left\langle A^{2}\right\rangle \simeq\frac{T\left(2D\right)^{3}}{\mu^{4}}\kappa_{2}
\ee
with some (dimensionless) coefficient $\kappa_2$.
Comparing Eqs.~\eqref{varVsTwithD1} and \eqref{varVsTgeneralmuD}, we find that $\kappa_2 = 5/4$, leading to Eq.~\eqref{varVsTDHalf} of the main text.

\section{Perturbative DV}
\label{app:DV}

\renewcommand{\theequation}{B\arabic{equation}}
\setcounter{equation}{0}

 Since $\psi(x)$ is only determined up to an overall normalization factor, we arbitrarily choose $C_1=1$ in Eq.~\eqref{psisol}. Then, the continuity condition at $x=0$ yields
\be
C_2=\frac{\text{Ai}\left(\frac{-2E+1}{2^{2/3}k^{2/3}}\right)}{\text{Bi}\left(\frac{-2E+1}{2^{2/3}k^{2/3}}\right)}\,.
\ee
Integrating the Schrödinger equation
\eqref{Schrodinger} over an infinitesimal interval around $x=0$, one obtains the condition
\be
0 = -\frac{1}{2}\left[\psi'\left(0^{+}\right)-\psi'\left(0^{-}\right)\right]-\psi\left(0\right) \, .
\ee
Inserting Eq.~\eqref{psisol} (with $C_1$ and $C_2$ as given above) into this condition, we find
\bea
\label{Eofk1}
0&=&-\frac{1}{2}\left[2^{1/3}k^{1/3}\text{Ai}'\left(\frac{-2E+1}{2^{2/3}k^{2/3}}\right)-2^{1/3}k^{1/3}\frac{\text{Ai}\left(\frac{-2E+1}{2^{2/3}k^{2/3}}\right)}{\text{Bi}\left(\frac{-2E+1}{2^{2/3}k^{2/3}}\right)}\text{Bi}'\left(\frac{-2E+1}{2^{2/3}k^{2/3}}\right)\right]-\text{Ai}\left(\frac{-2E+1}{2^{2/3}k^{2/3}}\right) \nn\\
&=&\frac{k^{1/3}}{2^{2/3}\pi\text{Bi}\left(\frac{1-2E}{2^{2/3}k^{2/3}}\right)}-\text{Ai}\left(\frac{1-2E}{2^{2/3}k^{2/3}}\right) \, ,
\eea
where we used that the Wronskian of the Airy functions is
\be
\text{Ai}\left(z\right)\text{Bi}'\left(z\right)-\text{Ai}'\left(z\right)\text{Bi}\left(z\right)=\frac{1}{\pi}.
\ee
After slightly rearranging Eq.~\eqref{Eofk1}, one reaches Eq.~\eqref{Eofk2} of the main text.

\section{Properties of the rate function $f(y)$}
\label{app:fofy}

\renewcommand{\theequation}{C\arabic{equation}}
\setcounter{equation}{0}

Rate functions very similar to $f(y)$ that is given in Eq.~\eqref{fsol} have been encountered and analyzed in other contexts, see e.g. Refs.~\cite{Smith22OU, SmithMajumdar22}. However, for completeness we include here a brief analysis of the main properties of $f(y)$.

Let us denote the function that appears on the right-hand side of Eq.~\eqref{fsol} by
\be
F\left(y,z\right)=\frac{2^{3/2}\sqrt{z}}{3^{1/2}}+\frac{2}{5}\left(y-z\right)^{2}
\ee
so that
$f\left(y\right)=\min_{0\le z\le y}F\left(y,z\right)$.
We begin by finding the (local) minima of $F$ as a function of $z$. One local minimum is at $z=0$, which yields one branch of the rate function, $f_0(y) = 2y^{2}/5$.
One can look for other minima by searching for solutions $z=z_*$ to the equation
\be
\partial_{z}F\left(y,z\right)=\sqrt{\frac{2}{3z}}+\frac{4}{5}\left(z-y\right)=0 \, .
\ee
Solving this equation for $y$, we obtain
\be
\label{yofzstar}
y=z_{*}+\frac{5}{2\sqrt{6z_{*}}} \, ,
\ee
which may then be plugged back into Eq.~\eqref{fsol} to give a second branch of the rate function,
\be
\label{fofzstar}
f_1=\frac{2^{3/2}\sqrt{z_{*}}}{3^{1/2}}+\frac{5}{12z_{*}} \, .
\ee
Eqs.~\eqref{yofzstar} and \eqref{fofzstar} give $f_1(y)$ in a parametric form.

\begin{figure*}[ht]
\includegraphics[width=0.6\linewidth,clip=]{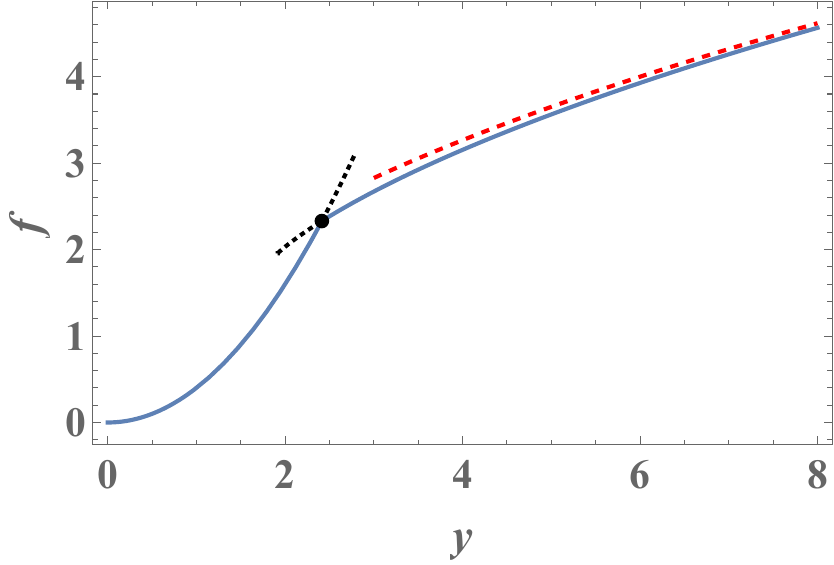}
\caption{Solid line: The rate function $f(y)$. The fat dot denotes the critical point $y=y_c$ at which $f(y)$ exhibits a corner singularity. The dashed line corresponds to the asymptotic behavior $f(y\gg1) \simeq 2^{3/2}\sqrt{y/3}$. The dotted lines correspond to the branches $f_0(y)$ and $f_1(y)$ in regimes of $y$ for which they are not optimal.}
\label{fig:fofy}
\end{figure*}

The true $f(y)$ is given by
\be
f\left(y\right)=\min\left\{ f_{0}\left(y\right),f_{1}\left(y\right)\right\}  \, .
\ee
One finds \cite{Smith22OU} that $f_1(y)$ exists at $y>y_\ell$ where
\be
y_\ell = \frac{15^{2/3}}{2^{5/3}} = 1.91577\dots,
\ee
and that $f_1(y) < f_0(y)$ at $y$ that is above a critical value
\be
y_{c}=\frac{15^{2/3}}{2^{4/3}}=2.41372\dots.
\ee
At $y<y_c$, $f(y) = f_0(y) =2y^{2}/5$ is exactly parabolic.
At $y=y_c$, the graphs of the functions $f_0(y)$ and $f_1(y)$ cross each other. As a result, $f(y)$ exhibits a corner singularity there.
The limit $y \gg 1$ corresponds to $z_* \gg 1$. In the leading order, one simply finds $z_* \simeq y$ so $f(y) \simeq 2^{3/2}\sqrt{y/3}$.
 $f(y)$ is plotted, along with its asymptotic behaviors, in Fig.~\ref{fig:fofy}.

\section{Absolute area}
\label{app:absoluteArea}

\renewcommand{\theequation}{D\arabic{equation}}
\setcounter{equation}{0}

In Ref.~\cite{BH20}, a transcendental equation whose solution gives $\lambda(k)$ at $k<0$ was obtained. In our rescaled variables, this equation reads
\be
\label{transcendentalB}
\left(-2k\right)^{1/3}\text{Ai}'\left[\frac{2\lambda+1}{\left(-2k\right)^{2/3}}\right]+\text{Ai}\left[\frac{2\lambda+1}{\left(-2k\right)^{2/3}}\right]=0 \, .
\ee
Instead of solving this equation numerically (as in \cite{BH20}), it is more convenient for our purposes to use it to obtain $\lambda(k)$ in a parametric form.
Denoting the argument of the Airy functions by
$\eta=\left(2\lambda+1\right)/\left(-2k\right)^{2/3}$,
Eq.~\eqref{transcendentalB} becomes
$\left(-2k\right)^{1/3}\text{Ai}'\left(\eta\right)+\text{Ai}\left(\eta\right)=0$, from which one obtains
\be
\label{kOfEtaB}
k\left(\eta\right)=\frac{1}{2}\left[\frac{\text{Ai}\left(\eta\right)}{\text{Ai}'\left(\eta\right)}\right]^{3} \, .
\ee
Plugging this back into the definition of $\eta$, we solve for $\lambda$ to obtain
\be
\label{lambdaOfEtaB}
\lambda\left(\eta\right)=\frac{\left[\text{Ai}\left(\eta\right)/\text{Ai}'\left(\eta\right)\right]^{2}\eta-1}{2} \, .
\ee
Eqs.~\eqref{kOfEtaB} and \eqref{lambdaOfEtaB} give $\lambda(k)$ in a parametric form (at $k<0$).

The limit $-k \ll 1$ corresponds to the limit $\eta \gg 1$. In this limit, Eqs.~\eqref{kOfEtaB} and \eqref{lambdaOfEtaB} become
\bea
k\left(\eta\right)&\simeq&-\frac{1}{2\eta^{3/2}}+\frac{3}{8\eta^{3}}-\frac{27}{64\eta^{9/2}}+\frac{85}{128\eta^{6}}-\frac{5775}{4096\eta^{15/2}}\,,\\
\lambda\left(\eta\right)&\simeq&-\frac{1}{4\eta^{3/2}}+\frac{1}{4\eta^{3}}-\frac{49}{128\eta^{9/2}}+\frac{105}{128\eta^{6}}-\frac{19019}{8192\eta^{15/2}}\,,
\eea
respectively, from which one extracts the asymptotic behavior
\be
\lambda\left(k\right)\simeq\frac{k}{2}+\frac{k^{2}}{4}+\frac{5k^{3}}{8}+\frac{11k^{4}}{4}+\frac{539k^{5}}{32} \, .
\ee
Using Eq.~\eqref{kappanDef} one can again obtain the coefficients describing the linear growth of cumulants in time. In particular, we have hear $\kappa_1 = 1/2$ and $\kappa_2 = 1/2$, yielding the mean $\bar{B} \simeq T/2$ and variance $\text{Var}\left(B\right)\simeq T/2$ of $B$ respectively, as given in the main text.

Incidentally, the rate function $\tilde{I}(b)$ may also be obtained (at $0<b<1/2$) in a parametric form, by taking the Legendre transform and using the chain rule:
\bea
b&=&\frac{d\lambda}{dk}=\frac{d\lambda/d\eta}{dk/d\eta}=\frac{\text{Ai}'(\eta)\left[-2\eta^{2}\text{Ai}(\eta)^{2}+2\eta\text{Ai}'(\eta)^{2}+\text{Ai}(\eta)\text{Ai}'(\eta)\right]}{3\text{Ai}(\eta)\left[\text{Ai}'(\eta)^{2}-\eta\text{Ai}(\eta)^{2}\right]}\, ,\\
\tilde{I}\left(b\right)&=&kb-\lambda\left(k\right)=-\frac{\eta^{2}\text{Ai}(\eta)^{4}+3\text{Ai}'(\eta)^{4}+\text{Ai}(\eta)^{3}\text{Ai}'(\eta)-4\eta\text{Ai}(\eta)^{2}\text{Ai}'(\eta)^{2}}{6\eta\text{Ai}(\eta)^{2}\text{Ai}'(\eta)^{2}-6\text{Ai}'(\eta)^{4}}\, .\nn\\
\eea





\end{document}